\documentclass[twocolumn,prl,reprint,secnumarabic,amssymb,amsmath,nobibnotes,aps,showpacs]{revtex4-1}
\pdfoutput=1
\usepackage{graphics}
\usepackage{graphicx}
\usepackage{amssymb}
\usepackage{braket}

\usepackage[usenames,dvipsnames]{color}

\begin{document}

\title{Magnetically-Trapped Molecules Efficiently Loaded from a Molecular MOT}

\author{D.~J. McCarron} 
\thanks{Present address: Department of Physics, University of Connecticut, 2152 Hillside Road, Unit 3046, Storrs, CT 06269-3046. Corresponding author: daniel.mccarron@uconn.edu.}
\affiliation{Department of Physics, Yale University, PO Box 208120, New Haven, Connecticut 06520, USA}
\author{M.~H. Steinecker}
\affiliation{Department of Physics, Yale University, PO Box 208120, New Haven, Connecticut 06520, USA}
\author{Y. Zhu}
\affiliation{Department of Physics, Yale University, PO Box 208120, New Haven, Connecticut 06520, USA}
\author{D. DeMille}
\affiliation{Department of Physics, Yale University, PO Box 208120, New Haven, Connecticut 06520, USA}

\begin{abstract}
We describe experiments demonstrating efficient transfer of molecules from a magneto-optical trap (MOT) into a conservative magnetic quadrupole trap. Our scheme begins with a blue-detuned optical molasses to cool SrF molecules to $\sim\!50$~$\mu$K. Next, we optically pump the molecules into a strongly-trapped sublevel. This two-step process reliably transfers $64\%$ of the molecules initially trapped in the MOT into the magnetic trap, comparable to similar atomic experiments. Once loaded, the magnetic trap is compressed by increasing the magnetic field gradient. Finally, we demonstrate a magnetic trap lifetime of over $1$~s. This opens a promising new path to the study of ultracold molecular collisions, and potentially the production of quantum-degenerate molecular gases.
\end{abstract}

$\pacs{}$

\maketitle

In the last decade, there has been tremendous progress in the production of samples of ultracold polar molecules, beginning with techniques to assemble molecules from pre-cooled alkali atoms \cite{Ni2008,Danzl2010,Aikawa2010,Shimasaki2015}. These efforts have been motivated by the many proposed applications for ultracold molecules in precision measurement \cite{Hunter2012,Tarbutt2013,Altuntas2017}, quantum information \cite{DeMille2002,Rabl2006,Andre2006} and quantum simulation \cite{Micheli2006}, and ultracold chemistry \cite{Krems2008}. For many of these applications, it is critical to use a sample with high phase-space density \cite{Carr2009}.

Recently, techniques for direct laser cooling \cite{Shuman2010} and magneto-optical trapping of molecules \cite{Barry2014,Norrgard2016,Truppe2017} have been developing rapidly. Methods based on these starting points are being pursued as an alternate route to produce ultracold gases of polar molecules. There have been steady improvements in the number density and temperature achieved via these methods \cite{Truppe2017,Steinecker2016,Anderegg2017}. Nonetheless, the phase-space densities accessible via direct molecular cooling and trapping remain quite modest compared to those of their atomic counterparts. The success of sympathetic and evaporative cooling techniques for increasing the phase-space density of atomic samples \cite{Bloch2001,Ketterle1996} provides inspiration for the pursuit of similar techniques using ultracold molecules. The first step towards implementing these techniques in molecular samples is to load molecules into a conservative trap.

The first magnetic trapping of a molecule was reported in Ref. \cite{Weinstein1998}, using superconducting coils in a cryogenic buffer gas cell. Since then, several other experiments have observed magnetic \cite{Lu2014}, electrostatic \cite{Zeppenfeld2012}, or mixed magnetic/electric \cite{Sawyer2007,Reens2017} conservative trapping of directly-cooled molecules. However, in these experiments the temperature of the molecular cloud was always $\gtrsim 400$~$\mu$K, and typically $\gg \! 1$~mK.  Here, we report magnetic trapping of molecules laser-cooled to temperatures well below $100$~$\mu$K.

The magnetic quadrupole trap (MQT) combines substantial trap depth ($\gtrsim \! 10$~mK using normally conducting coils) with large trapping volume ($\gtrsim1$ cm$^{3}$) that match well to laser-cooled atomic and molecular clouds. The linear potential from an MQT also provides tight confinement and affords a simple experimental implementation. However, there are drawbacks to working with an MQT. One is an inability to confine the absolute internal ground state, so trapped samples can suffer losses due to two-body inelastic collisions. Another is vulnerability to Majorana spin-flip losses, which occur for sufficiently low-temperature samples \cite{Petrich1995}. Given the densities ($\lesssim\!10^{6}$~cm$^{-3})$ and temperatures ($\sim50$~$\mu$K) presently accessible in molecular laser cooling and trapping experiments, neither of these mechanisms is expected to be limiting. This makes the MQT an excellent choice to realize the efficient transfer of molecules from a magneto-optical trap (MOT) into a conservative trap. Magnetic trapping of ultracold molecules has hence been a long-standing goal, as an attractive starting point for future studies of sympathetic cooling and ultracold chemistry \cite{Carr2009}.

In this Letter, we demonstrate and characterize the transfer of SrF molecules from a radio-frequency (RF) MOT into an MQT. We apply sub-Doppler cooling followed by optical pumping (OP) to reliably transfer $64(2)\%$ of the molecules initially trapped in the RF MOT. Once in the MQT, we increase the field gradient to compress the trapped cloud. Finally, we demonstrate a trap lifetime in excess of $1$~s.


Our experimental setup for the RF MOT has been described in detail elsewhere \cite{Barry2014,McCarron2015,Norrgard2016,Steinecker2016}. In brief, pulses of SrF molecules are produced using a cryogenic buffer gas beam source \cite{Barry2011,Hutzler2012} and slowed with a ``white light'' scheme employing lasers $\mathcal{L}_{00}$, $\mathcal{L}_{10}$ and $\mathcal{L}_{21}$ \cite{Barry2012}. Here, $\mathcal{L}_{vv'}$ denotes a laser tuned to the $X\,^{2}\Sigma^+ \ket{v,N^P=1^-}\rightarrow A\,^{2}\Pi_{1/2} \ket{v',J^{\prime P'}=1/2^+}$ transition, where $v$ is the vibrational quantum number, $N$ is the angular momentum excluding spin, $J$ is the angular momentum excluding nuclear spin, $P$ is the parity, and prime indicates the excited state. Slow molecules are captured in the RF MOT, which includes the additional repump laser $\mathcal{L}_{32}$ \cite{Norrgard2016}. The RF MOT coils produce undesired electric fields that mix the $J^{\prime P'}=1/2^-$ and $J^{\prime P'}=1/2^+$ excited states and cause branching into the $N^P=0^+,2^+$ ground states. To return these molecules to the optical cycle, a repump laser, $\mathcal{L}^{N=2}_{00}$ (driving $\ket{v=0,N^P=2^+}\rightarrow \ket{v'=0,J^{\prime P'}=1/2^-}$), and microwaves (driving $\ket{v=0,N^P = 0^+; J = 1/2; F = 0, 1}\leftrightarrow \ket{v=0,N^P = 1^-; J = 1/2; F = 1, 0}$) are applied in the RF MOT \cite{Norrgard2016}. RF sidebands are added to each of the MOT lasers to address the resolved spin-rotation/hyperfine (SR/HF) structure in the SrF ground state. Molecular laser cooling uses type-II cycling transitions ($F\rightarrow F' = F$ or $F-1$, where $F$ is the total angular momentum) \cite{Barry2014}, where there exist ground-state sublevels not optically coupled to the excited state (i.e., dark states) for any fixed laser polarization \cite{DiRosa2004,Stuhl2008}. The RF MOT destabilizes these dark states by rapidly and synchronously reversing the trapping laser polarizations and the MOT magnetic field gradient \cite{Hummon2013}. More details on the RF MOT, including one additional repump laser, are given in the Supplemental Materials. An acousto-optic modulator (AOM) allows the MOT trapping laser intensity to be rapidly changed from the maximum value. A mechanical shutter allows all of the MOT light to be extinguished. Trapped molecules are detected by imaging laser-induced fluorescence (LIF) from the cycling transition with a CCD camera.

We define the time of the ablation laser pulse that initiates a molecular beam pulse as $t=0$. To load the RF MOT, we apply the slowing lasers to the molecular beam from $t=0$ to $t=35$~ms and apply the trapping lasers at their maximum intensity from $t=0$ until $t=67$~ms. Once captured, the molecules are cooled and compressed by simultaneously reducing the trapping laser intensity over $50$~ms and increasing the amplitude of the RF $B$-field 
gradient by a factor of $2$ \cite{Steinecker2016}. The laser intensity and gradient amplitude are then held at their new values for $20$~ms. 

Next, we apply a sub-Doppler cooling stage. This allows us to choose the final trap laser intensity during MOT compression so as to produce clouds with maximum density (as opposed to minimum temperature or maximum phase-space density, as in our previous work \cite{Steinecker2016}). Here, the trapping laser intensity is reduced to $5$\% of the maximum value to produce clouds at $1.1(1)$~mK, with density $1.8\times$ greater than when optimized for maximum phase-space density. 

The sub-Doppler cooling is achieved using a blue-detuned molasses, similar to that recently demonstrated for CaF molecules \cite{Truppe2017}. At $t=137$~ms, the RF MOT magnetic field gradient, polarization switching, and resonant microwaves are switched off. The trapping light is extinguished for $0.1$~ms as the trapping laser frequencies are jumped up by $+4.2\Gamma$ from the optimal trapping values (where $\Gamma=2\pi\times 6.6$~MHz is the natural linewidth); this produces a blue-detuned molasses with detuning $\approx \! +2.8\Gamma$ for each SR/HF level. The trapping light is then restored and applied to the molecules for $1.8$~ms. All repump laser light is also present during this molasses stage. During the molasses, three orthogonal shim coils, centered on the trapping region, are rapidly switched on (within $\approx \! 1$~ms) to produce small, tunable magnetic fields. After the molasses, all MOT light (trapping and repump) is extinguished and the trapping laser frequencies are returned to the original values used in the RF MOT for subsequent fluorescence detection of the molecular cloud.

\begin{figure}[t b]
\centering
\includegraphics[clip, scale=0.23]{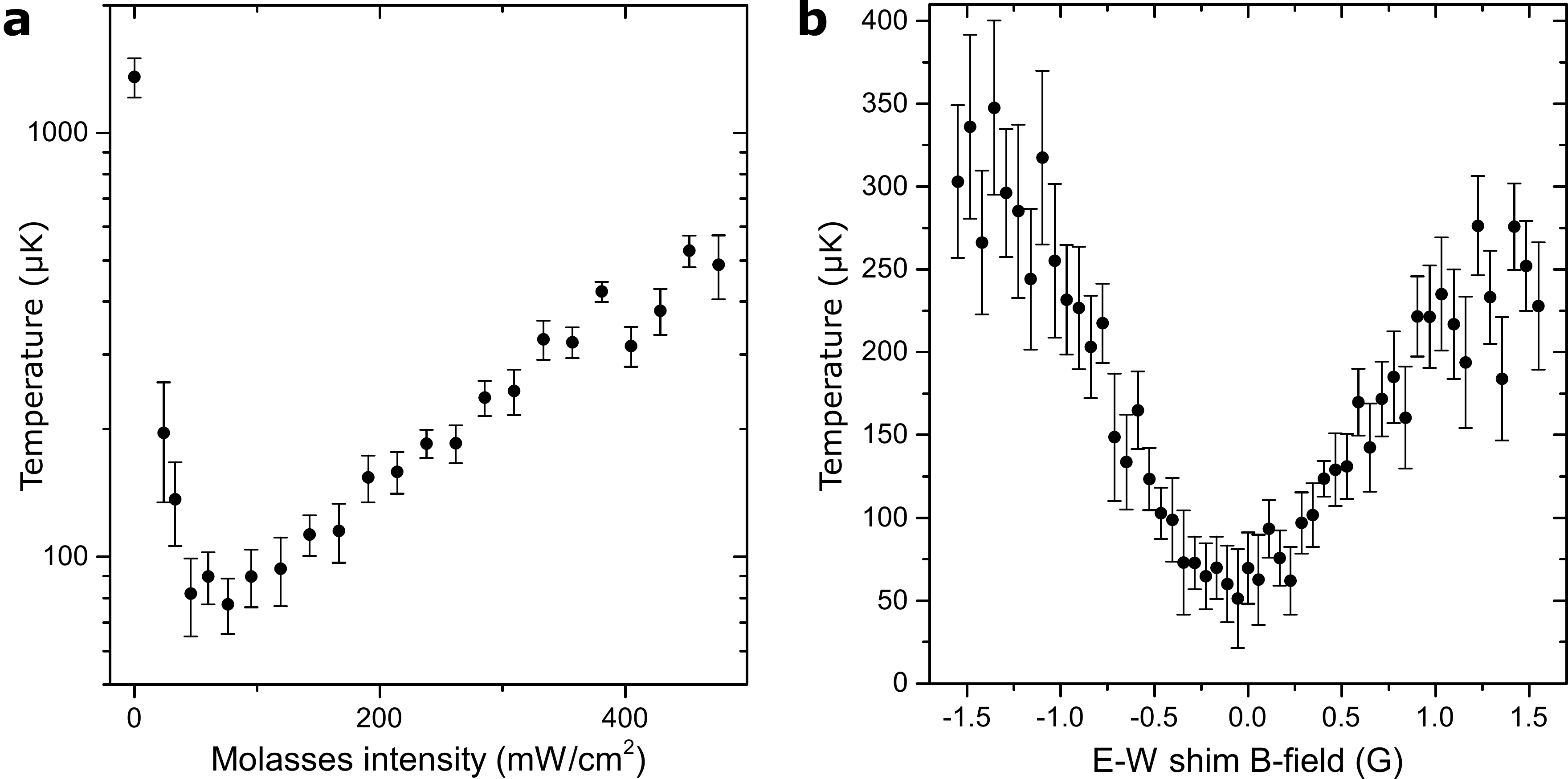}
\caption{Sub-Doppler Cooling: \textbf{a} Molecular temperature vs. molasses laser intensity. The $0$~mW/cm$^{2}$ point indicates the initial cloud temperature, prior to the molasses. \textbf{b} Molecular temperature after molasses vs. east-west shim coil $B$-field. Temperature dependence on north-south and up-down shim coil $B$-fields is similar (see Supplemental Materials). } \label{fig:temperature}
\end{figure}

At this and all later stages in the experiment, $\ket{N^P=0^+}$ state 
molecules populated by microwaves during the MOT stage (see Supplemental Materials) no longer play a role in the signals (they are not addressed by the detection lasers). Assuming all 28 ground-state sublevels linked by lasers or microwaves are equally populated in the MOT \cite{Tarbutt2013}, we expect $\ket{N^P=1^-}$ state molecules to comprise $\approx \! 24/28 = 86$\% of the molecules in the MOT.  All discussion going forward refers to this $\ket{N^P=1^-}$ state population.

Molasses settings were optimized to produce clouds with minimum temperature, as determined by time-of-flight (TOF) measurements (Fig. 1). We find that optimal sub-Doppler cooling in SrF requires a magnetic field with a specific magnitude and orientation, and a total trapping laser intensity of $\approx \! 60$~mW/cm$^{2}$. (All laser intensities reported are averages over the $1/e^2$ area of the laser beams.) Our empirically-determined optimum field approximately cancels Earth's field in our laboratory. The optimized molasses cools molecules to temperatures as low as $T_\text{mol}=50(10)~\mu$K, with negligible loss in number. This temperature is a factor of $\approx\!3$ smaller than the Doppler temperature for this transition in SrF, $T_{\rm{D}}=\hbar\Gamma/(2k_{\rm{B}})=160$~$\mu$K, and a factor of $\approx\! 5$ smaller than our previous lowest temperature \cite{Steinecker2016}.

Molecules are next optically pumped towards the $\ket{N^P = 1^-, J = 3/2, F = 2, m_{F}=+2}$ stretched-state Zeeman sublevel, by driving $\sigma^{+}$ and $\pi$ transitions on all SR/HF lines of the $\ket{v=0,N^P=1^-} \rightarrow \ket{v'=0,J^{\prime P'}=1/2^+}$ resonance. This stretched state provides the maximum magnetic confinement available in $X\,^{2}\Sigma^{+}\ket{v=0,N^P=1^-}$. (Approximately the same confinement is achieved in one additional state, $\ket{N^P = 1^-, J = 3/2, F = 1, m_{F}=+1}$.) To achieve the OP, a circularly-polarized laser beam, with intensity $\sim\!30$~mW/cm$^2$, is applied from $t=140$~ms, and the shim coil currents are rapidly set to provide a $\approx \! 3$~G quantizing field at a small angle to the propagation axis of the laser beam. The laser, with RF sidebands to address the relevant SR/HF structure, is tuned to the field-free resonance (defined as the carrier frequency that gives maximum LIF when the laser beam is applied transverse to the molecular beam). 

\begin{figure}[t b]
\centering
\includegraphics[clip, scale=0.35]{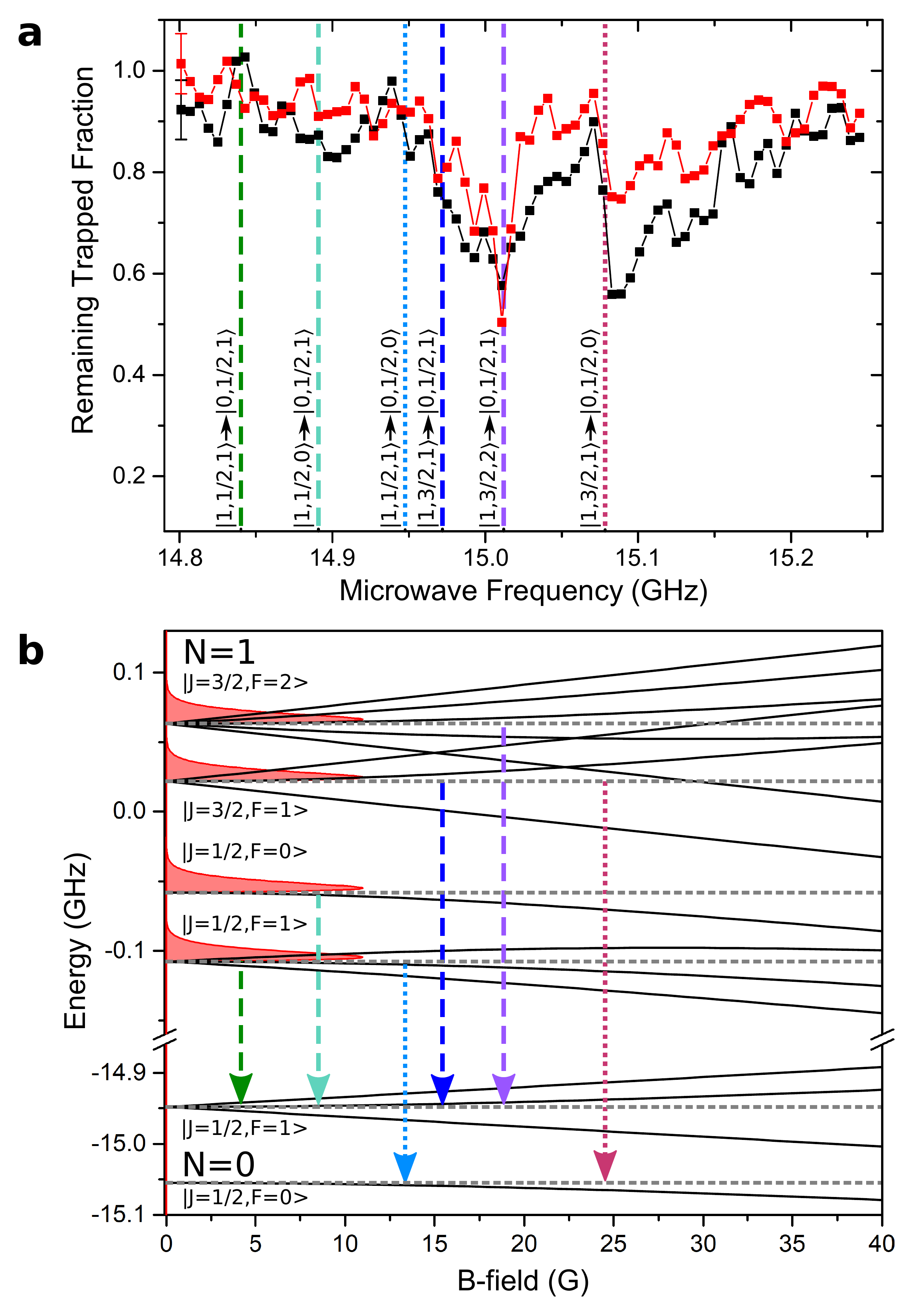}
\caption{(color online) Effect of OP: \textbf{a} Microwave spectroscopy in the compressed MQT, after loading without (black) and with (red) OP. Dashed (dotted) lines mark the field-free frequencies of microwave transitions to $\ket{v=0, N^P=0^+, J=1/2, F=1}$ (to $\ket{v=0, N^P=0^+, J=1/2, F=0}$), as illustrated in plot (b). The first data points show typical standard errors. State labels are shown with an abbreviated notation $\ket{N, J, F}$. \textbf{b} Breit-Rabi diagram for the $X^{2}\Sigma^{+}$ $\ket{v=0; N^P = 1^-; J = 1/2, 3/2}$ and $\ket{v=0, N^P = 0^+, J = 1/2}$ states. Vertical arrows mark the six available microwave transitions, with their horizontal positions aligned below the relevant frequencies in plot (a). Red shaded regions show the energy spread for molecules in each trapped hyperfine state, assuming a Boltzmann distribution at $240~\mu$K.
} \label{fig:OP}
\end{figure}

At $t=142$~ms, the OP light and quantizing magnetic field are switched off, and the MQT is rapidly switched on in $250$~$\mu$s, to an axial gradient of $32$~G/cm (oriented vertically). (The levitation gradient for the $\ket{F = 2, m_{F}=+2}$ state is $19~$G/cm.) The gradient can then be increased to a value as large as $140$~G/cm over the next $100$~ms to spatially compress the trapped cloud. The $3.6$~mH inductance of our MQT coils makes rapid changes in current technically challenging. We use driving circuitry substantially based on the designs described in Ref. \cite{Aubin2005} and discussed in more detail in the Supplemental Materials. 

To probe properties of the magnetically trapped molecules, including molecular cloud number and temperature, we use our standard TOF fluorescence imaging  \cite{Steinecker2016}. The trapping field is switched off rapidly, using another custom circuit (see Supplemental Materials).   The current in the trapping coils is fully switched off within $1$~ms, during which time the molecular cloud expands negligibly.   We apply all MOT lasers at their full intensities (but without the RF MOT $B$-field gradient or the microwaves linking the $\ket{N^P=0^+}$ and $\ket{N^P=1^-}$ states) to image the molecular cloud. We probe only after a time of flight of $\ge3$~ms (when eddy currents from the field switch-off have decayed to a negligible level).

When the trap is maintained approximately at its switch-on gradient for the duration of trapping (referred to as the low-gradient case), so that only molecules in the most strongly-trapped states are retained, we load $\approx\!40$\% of the molecules from the $\ket{N^P=1^-}$ state in the MOT into the MQT. In this low-gradient case, only slight heating (to $T_\text{trap}\approx90$~$\mu$K) is expected based on an analytic expression for the energy imparted by the switch-on of the MQT \cite{Ram2014}; we confirm this expectation in numerical simulations including the effect of gravity. Experimentally, we detect no substantial heating of the molecules by the trap. However, our signal-to-noise ratio in the low-gradient case is poor, leading to considerable uncertainty in the measured value of $T_\text{trap}$.

In the linear potential of an MQT, we expect the spatial distribution of molecules to deviate significantly from a Gaussian. This problem is especially severe in the low-gradient case (as opposed to when compression is applied), where the axial confining force is comparable to the force of gravity. Evidence of this distortion is apparent in our TOF images, at early times, in the low-gradient case. At present, we are unable to produce an analytic model of the spatial distribution in the MQT, and consequently have no suitable functional form with which to fit our data. Hence, we do not report quantitative measurements of the density in the MQT, but rather only of temperatures.

When compression is applied in the MQT, we load $75(2)\%$ of $\ket{N^P=1^-}$ state molecules from the MOT into the MQT, a factor of $\approx \! 2$ increase compared to the low-gradient case. This indicates that, even with OP applied, a substantial fraction of molecules, $\approx \! 50$\%, remain in relatively weakly-trapped states. (Although the switch-on gradient is sufficient to levitate and confine our target state and the $\ket{N^P=1^-, J=3/2, F=1,m_F=+1}$ state, other trappable states including $\ket{N^P=1^-, J=3/2, F=2, m_{F}=+1}$ and $\ket{N^P=1^-, J=3/2, F=1, m_{F}=0}$ are below the levitation threshold here. If the gradient increases sufficiently rapidly for compression, some molecules in the latter states are nonetheless trapped, whereas none can be trapped in the low-gradient case.) This loading efficiency, with compression, molasses, and OP applied, is comparable to efficiencies realized in experiments with ultracold atomic gases \cite{vanderStam2007}. 

To crudely estimate the expected temperature in the compressed MQT, we consider only molecules in the most strongly-trapped states. For these molecules, again, we expect an initial temperature $\approx \! 90$~$\mu$K. For adiabatic compression in an MQT, we expect $T_\text{trap} \propto G^{2/3}$, where $G$ is the final gradient \cite{Ketterle1999}. Under our conditions, the gradient increases by a factor of $\approx\! 4$ during the compression, corresponding to a factor of $\approx \! 2.5$ increase in $T_\text{trap}$. Hence we crudely expect $T_\text{trap}\approx 230$~$\mu$K after compression. To produce a more realistic estimate, we perform a numerical simulation for an ensemble with $50$\% of molecules in states with half the magnetic moment of the strongly-trapped states. Applying our standard data analysis to clouds from this simulation yields the apparent value $T_\text{trap}\approx 250$~$\mu$K. In the compressed MQT, we measure a molecular temperature of $\approx\!240$~$\mu$K, in good agreement with expectations from both these approaches. We note that the adiabaticity criterion ($d\omega/dt \ll \omega^2$, where $\omega$ is the trap frequency) is not well-satisfied early in the compression sequence applied here, which could lead to additional heating of the trapped sample (see Supplemental Materials). In our simulations, however, we see no clear evidence of such heating when comparing our compression ramp to a slower ramp.

We detect no increase in the in-trap temperature with OP versus without it. When only the optical molasses is applied, we transfer $43(3)\%$ of $\ket{N^P=1^-}$ state molecules to the compressed MQT, and when only the OP stage is applied we transfer $26(1)\%$. An OP duration of $2$~ms is found sufficient to maximize loading into the compressed MQT. Loading efficiency into the compressed MQT with OP is not strongly dependent on the OP laser intensity or the magnetic field magnitude; both can be decreased by factors of $\sim\!2$ and still yield similar results. Similarly, the magnetic field orientation can be varied by $\pm40^{\circ}$ with negligible change in loading efficiency. When neither the optical molasses nor OP stages are employed, we load just $21(1)\%$ of $\ket{N^P=1^-}$ state molecules. This highlights the importance of using both pre-cooling and OP to realize efficient transfer into the compressed MQT.

In this work, no attempt has been made to maximize the number of molecules captured in the RF MOT. Here we observe $N_\text{MQT}\approx1000$~molecules in the compressed trap. However, based on the loading efficiencies achieved here and the maximum number of molecules previously observed in our RF MOT,  $N_\text{MOT}^\text{max}\approx10^4$ \cite{Steinecker2016}, we expect that optimization could allow capture of up to $\approx \! 7\times10^3$~molecules in the MQT.

To probe the $\ket{F, m_{F}}$ state distribution within the compressed MQT, we use microwave spectroscopy. In particular, we monitor trap loss while driving transitions between various SR/HF levels of the $\ket{v=0,N^P=1^-}$ and $\ket{v=0,N^P=0^+}$ states, at $\sim \! 15$~GHz. This spectroscopy is performed for loading sequences both with and without OP and provides a direct measurement of how the $\ket{F,m_F}$ state distribution in the trap is modified by the OP (Fig. 2). Once the MQT is loaded and compressed, a single microwave frequency (with $\sim\!1$~W of power and broadened to $3~$MHz with uniform noise) is applied, starting at $t=242$~ms and for a duration of $200$~ms. The number of remaining $\ket{v=0,N^P=1^-}$ molecules is measured as a function of microwave frequency.

We detect no clear evidence of molecules trapped in $\ket{N^P=1^-, J=1/2, F=0,1}$ either with or without OP. This is expected, given that the only trappable sublevel ($\ket{N^P=1^-, J=1/2, F=1, m_{F}=-1}$) is weakly confining. Without OP, we detect loss features in the compressed MQT corresponding to molecules populating both $\ket{N^P=1^-, J=3/2, F=1}$ and $\ket{N^P=1^-, J=3/2, F=2}$. With OP, the number of molecules in $\ket{N^P=1^-, J=3/2, F=1}$ decreases by a factor of $\approx1.8$ as molecules are successfully pumped towards the target $\ket{N^P=1^-, J=3/2, F=2, m_{F}=+2}$ sublevel. Trap loss of molecules in the target $\ket{N^P=1^-, J=3/2, F=2, m_{F}=+2}$ sublevel occurs when these molecules are driven to the $\ket{N^P=0^+, J=1/2, F=1, m_{F}=+1}$ sublevel. Since both these sublevels have the same magnetic moment, all trapped molecules in the target sublevel are resonant for the same frequency, $15.01$~GHz, regardless of the energy spread associated with the Boltzmann distribution. This resonance produces the sharp loss feature in our data when OP is present. (Molecules in $\ket{N^P=0^+,m_F=+1}$ remain trapped, but are not detected.)  The decreased width, when OP is applied, of the loss feature that extends to the blue of this narrow resonance indicates that OP is at least somewhat effective in depopulating the $\ket{N^P=1^-, J=3/2, F=2, m_{F}=+1}$ sublevel, since this sublevel is linked via microwaves to states with differing magnetic moments.

\begin{figure}[b!]
\centering
\includegraphics[clip, scale=0.35]{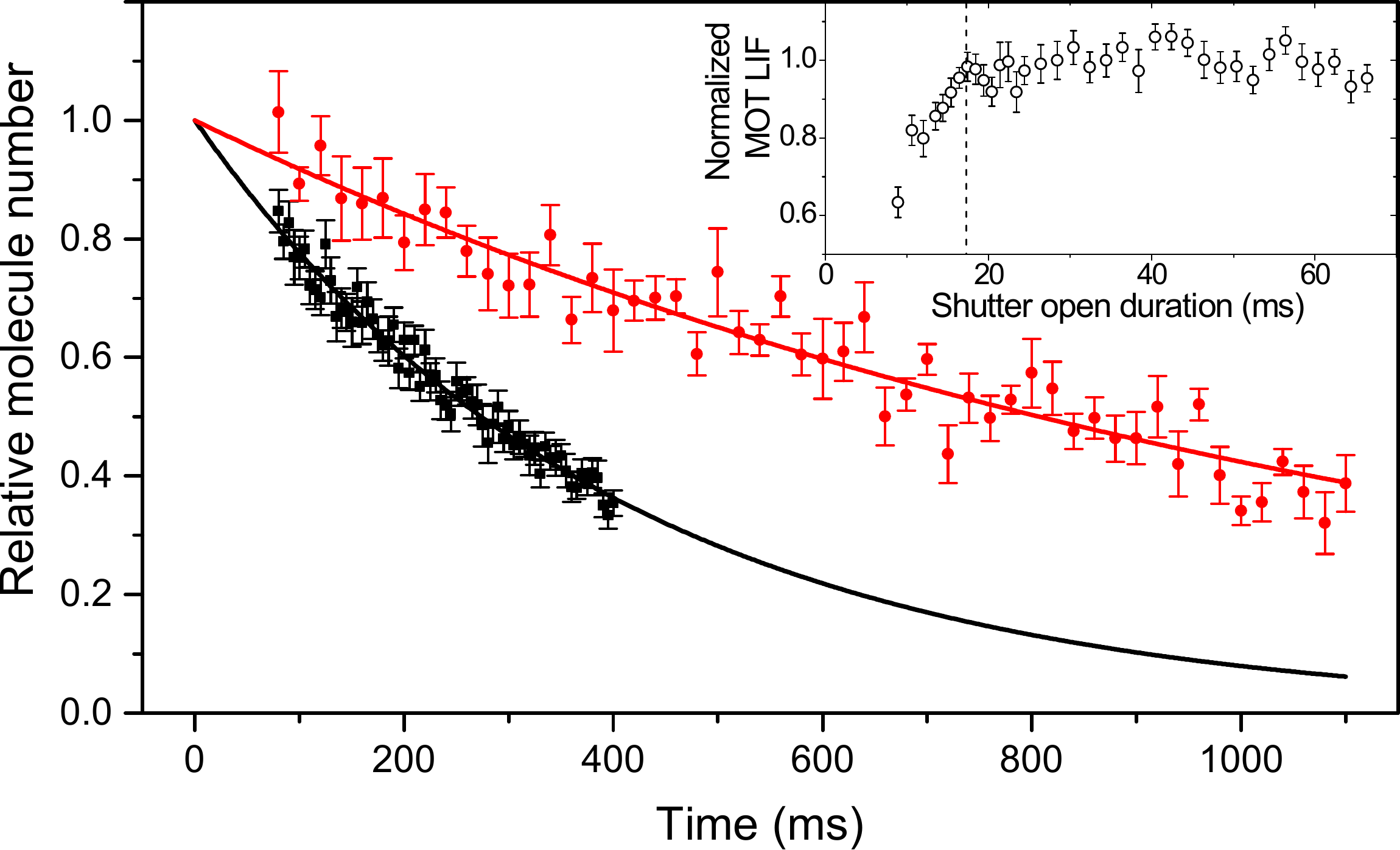}
\caption{(color online) Lifetime of molecules in the MQT. The lifetime with the in-vacuum shutter always open (black squares) is $395(9)$~ms. With the shutter open only from $t=0$ to $\Delta t=17$~ms, we measure an increased lifetime (red circles) of $1.17(5)$~ms.  Inset: MOT LIF signal vs. shutter open duration.  Setting $\Delta t = 17$~ms (dashed line) does not decrease the number of molecules initially loaded in the MOT. Error bars show the standard error of multiple measurements; lines are fits to an exponential.} \label{fig:MT}
\end{figure}

The MQT lifetime is measured by detecting LIF as a function of hold duration in the MQT. Initial measurements in the compressed MQT gave a lifetime of $395(9)$~ms, comparable to the maximum lifetime in our RF MOT \cite{Norrgard2016}, $370(40)$~ms under similar vacuum conditions. This indicates that the loss is dominated by collisions with background gases for both the MQT and the MOT.

To increase the MQT lifetime, an in-vacuum shutter was added to the molecular beam line, to reduce the helium gas load from the cryogenic source (Fig. 3). This shutter, located $1.2$~m upstream from the RF MOT center, is set to be fully open from $t=0$ to $\Delta t$. We find the initial number of molecules in the RF MOT is unchanged when $\Delta t \geq 17$~ms (Fig. 3 inset). With $\Delta t = 17$~ms, we measure an increased compressed MQT lifetime of $1.17(5)$~s while leaving $N_\text{MQT}$ unchanged. With the experiment repetition rates used here ($0.4$--$1.4$~Hz), this shutter duty cycle is observed to reduce the helium background pressure in the trapping region by a factor of $\sim \! 5$, down to below the total pressure of all other background gases ($\sim\! 5 \times 10^{-10}$~torr). 

At present, the in-trap molecule density is too low to detect inelastic molecule-molecule collisions, while the trap depth is large compared to the temperature for all relevant conditions. Therefore, the MQT lifetime is expected to be independent of the magnetic field gradient. To confirm this, we also measured the lifetime in the low-gradient case. Under this condition, we measure a lifetime of $1.04(9)$~s. This is, as expected, negligibly different from the lifetime in the fully compressed trap.

In summary, we have demonstrated efficient transfer of ultracold molecules from an RF MOT into an MQT. Our transfer scheme is similar to those employed in experiments using ultracold atoms and achieves a comparable transfer efficiency of $64(2)\%$ for all molecules in the RF MOT. Additional optimization of the OP by monitoring signals in the low-gradient MQT (rather than in the compressed MQT, as was done here) may increase the fraction of molecules pumped into the target state. An additional stage of OP prior to the molasses could recover the $\approx \! 14$\% of molecules left in the $\ket{N^P=0^+}$ state, raising the overall loading efficiency further. The demonstrated lifetime, in excess of $1$~s and limited only by background pressure, is likely sufficient to begin exploring atom-molecule collisions and the prospects for sympathetic cooling upon the addition of a co-trapped atomic sample \cite{Lim2015}. These investigations will naturally lead to studies of controlled ultracold chemistry and attempts to tune elastic and inelastic collision rates. In the longer term, the increasingly high phase-space density accessible in conservatively trapped samples of polar molecules holds enormous promise for future generations of precision measurement experiments \cite{Baron2014,Cairncross2017,Hunter2012}.

We thank M. Tarbutt for useful discussions on sub-Doppler cooling of molecules. The authors acknowledge support from ARO, ARO (MURI) and ONR. 

\bibliography{MTPaperBibliography}

\clearpage

\appendix \section{Supplemental Materials for \\ Magnetically-Trapped Molecules Efficiently Loaded from a Molecular MOT}

\begin{figure}[b]
\includegraphics[width=\columnwidth]{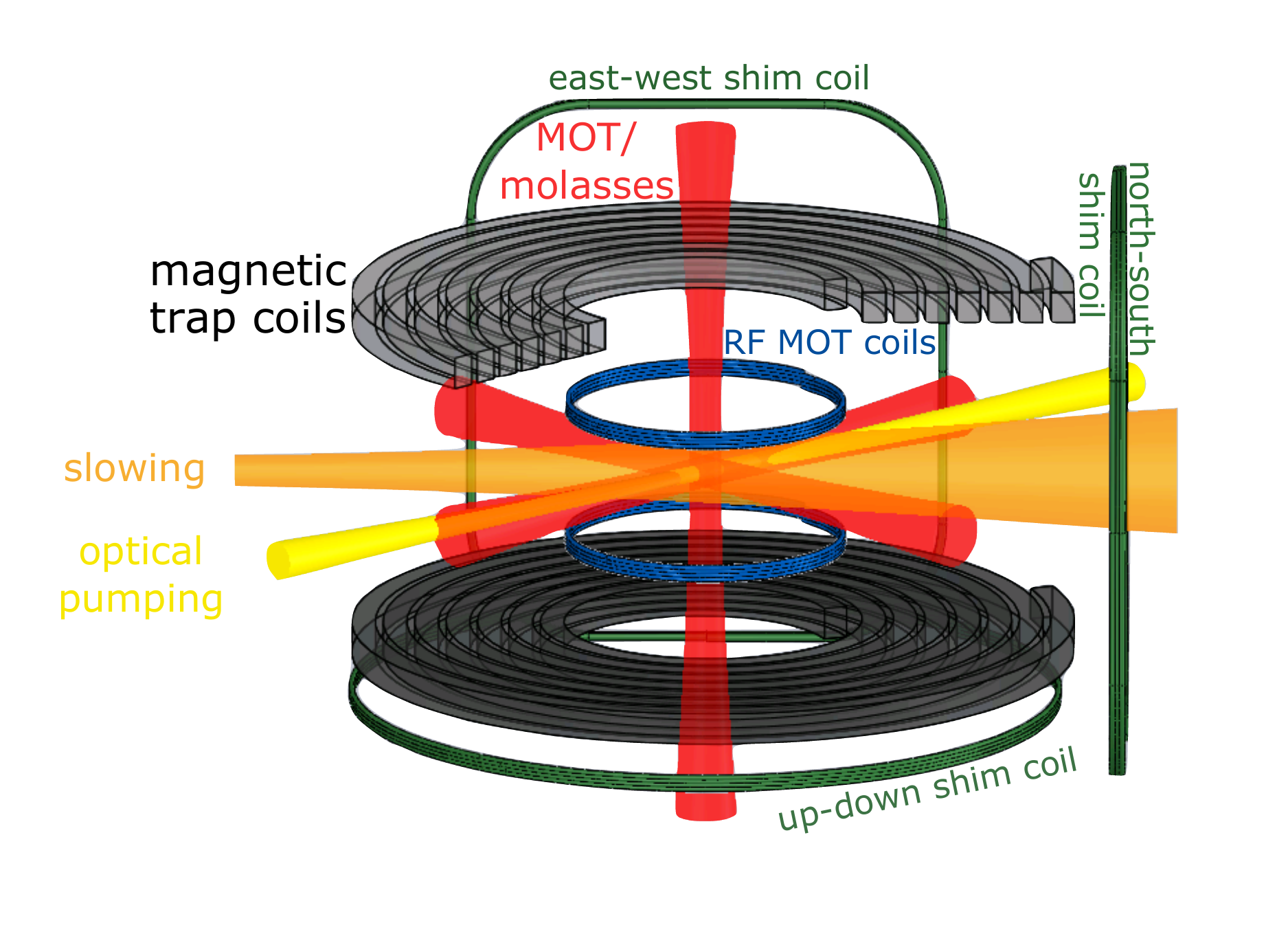}
\caption{(color online) Experimental setup. Molecules from a cryogenic buffer gas beam source (not shown) are slowed and trapped in the RF MOT. The RF MOT $B$-field gradient is switched off, and the molecules are further cooled in an optical molasses, using the MOT laser beams at reduced intensity and blue detuning. The MOT trapping lasers are then switched off, and an additional laser beam optically pumps the molecules into the most strongly-trapped Zeeman sublevel. All lasers are shuttered, and a DC magnetic field gradient is rapidly switched on to capture the molecules in the conservative MQT. During the molasses and optical pumping stages, three orthogonal shim coils provide the necessary $B$-fields for optimal performance.}
\label{fig:experiment}
\end{figure}

\begin{figure}[t b]
\includegraphics[width=\columnwidth]{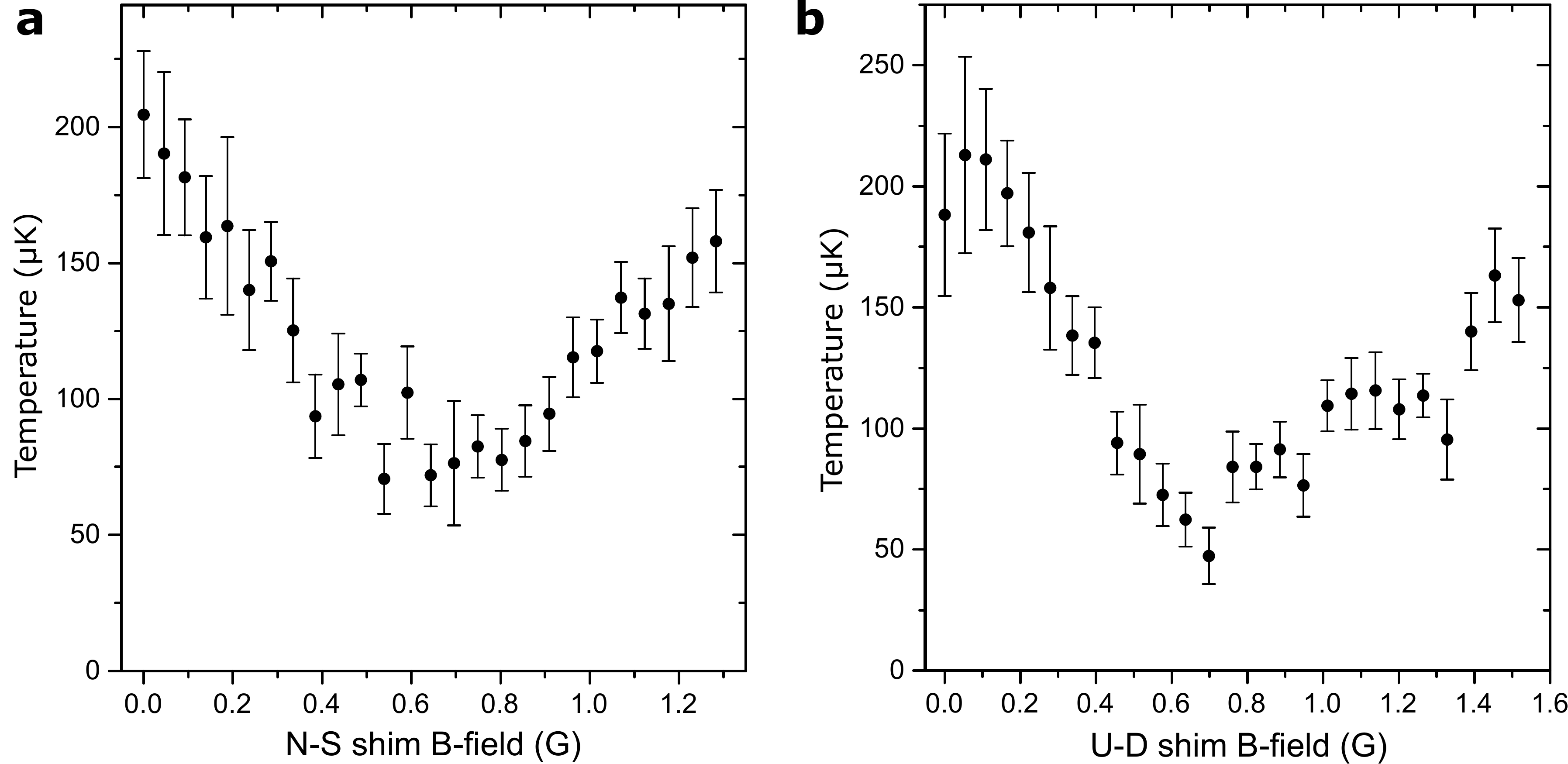}
\caption{Molecular temperature after molasses vs. $B$-field in \textbf{a} north-south and \textbf{b} up-down shim coils. These data show a qualitatively similar dependence of temperature on shim coil $B$-field along the north-south and up-down dimensions as seen in the east-west dimension (see Fig. 1 in the main text).}
\label{fig:molasses-secondhalf}
\end{figure}

The experimental setup for magnetic trapping of SrF molecules is shown as Fig. \ref{fig:experiment}. Additional details regarding the RF MOT and rapid current switching circuit for magnetic trapping, as well as a discussion of the adiabaticity of the MQT compression sequency, are given in the sections below. Figure \ref{fig:molasses-secondhalf} shows the dependence of the molecular temperature after the molasses stage on magnetic field in the north-south and up-down shim coils.

\section*{Additional details on RF MOT}

Our RF MOT uses a switching frequency of 1.23 MHz and a maximum root-mean-square (rms) axial $B$-field gradient of $18$~G/cm that requires a peak voltage of $3$~kV. This voltage produces an unwanted electric field in the RF MOT region that mixes the $\ket{J^{\prime P'}=1/2^-}$ and $\ket{J^{\prime P'}=1/2^+}$ excited states and leads to branching into the $N^P=0^+,2^+$ ground states. To combat this loss and return molecules to the optical cycling transition, a repump laser, $\mathcal{L}^{N=2}_{00}$, drives transitions $\ket{v=0,N^P=2^+}\rightarrow \ket{v'=0,J^{\prime P'}=1/2^-}$, and microwaves resonant with both $\ket{v=0,N^P = 0^+; J = 1/2; F = 0, 1}\leftrightarrow \ket{v=0,N^P = 1^-; J = 1/2; F = 1, 0}$ transitions are applied during the RF MOT stage [16]. In addition to returning molecules in $\ket{N^P=0^+}$ states to the optical cycle, the microwaves applied during the MOT stage also drive population from the $\ket{N^P=1^-}$ states to $\ket{N^P=0^+}$. Molecules that remain in $\ket{N^P=0^+}$ when the microwaves and MOT light are switched off are not addressed in the molasses or optical pumping stages, nor are they addressed by the imaging lasers. The RF MOT light also contains the repump laser $\mathcal{L}^{N=3}_{00}$ (driving $\ket{v=0,N^P=3^-}\rightarrow \ket{v'=0,J^{\prime P'}=3/2^+}$) to address decays into $\ket{v=0, N^P=3^-}$ due to accidental excitation of the $\mbox{$J^{\prime P'}=3/2^+$}$ states [28].

RF sidebands are added to each of the MOT lasers to address the resolved spin-rotation/hyperfine (SR/HF) structure in the SrF ground state [15]. As described in Ref. [15], one ground-state SR/HF level requires a different polarization for trapping, so an additional single-frequency laser, $\mathcal{L}^{\dagger}_{00}$, with polarization opposite to $\mathcal{L}_{00}$, is tuned closer to resonance with this level than the closest $\mathcal{L}_{00}$ sideband in order to approximate the optimum polarization scheme. The maximum powers of the $\mathcal{L}_{00}$ and $\mathcal{L}_{00}^\dagger$ trapping lasers are $110$~mW and $30$~mW, respectively, corresponding to intensities $430$~mW/cm$^2$ and $120$~mW/cm$^2$, respectively (including all six passes of the beams through the trapping region).

\section*{Rapid switching circuit for magnetic trapping coils}

\begin{figure*}[t b]
    \centering
    \includegraphics[width=\textwidth]{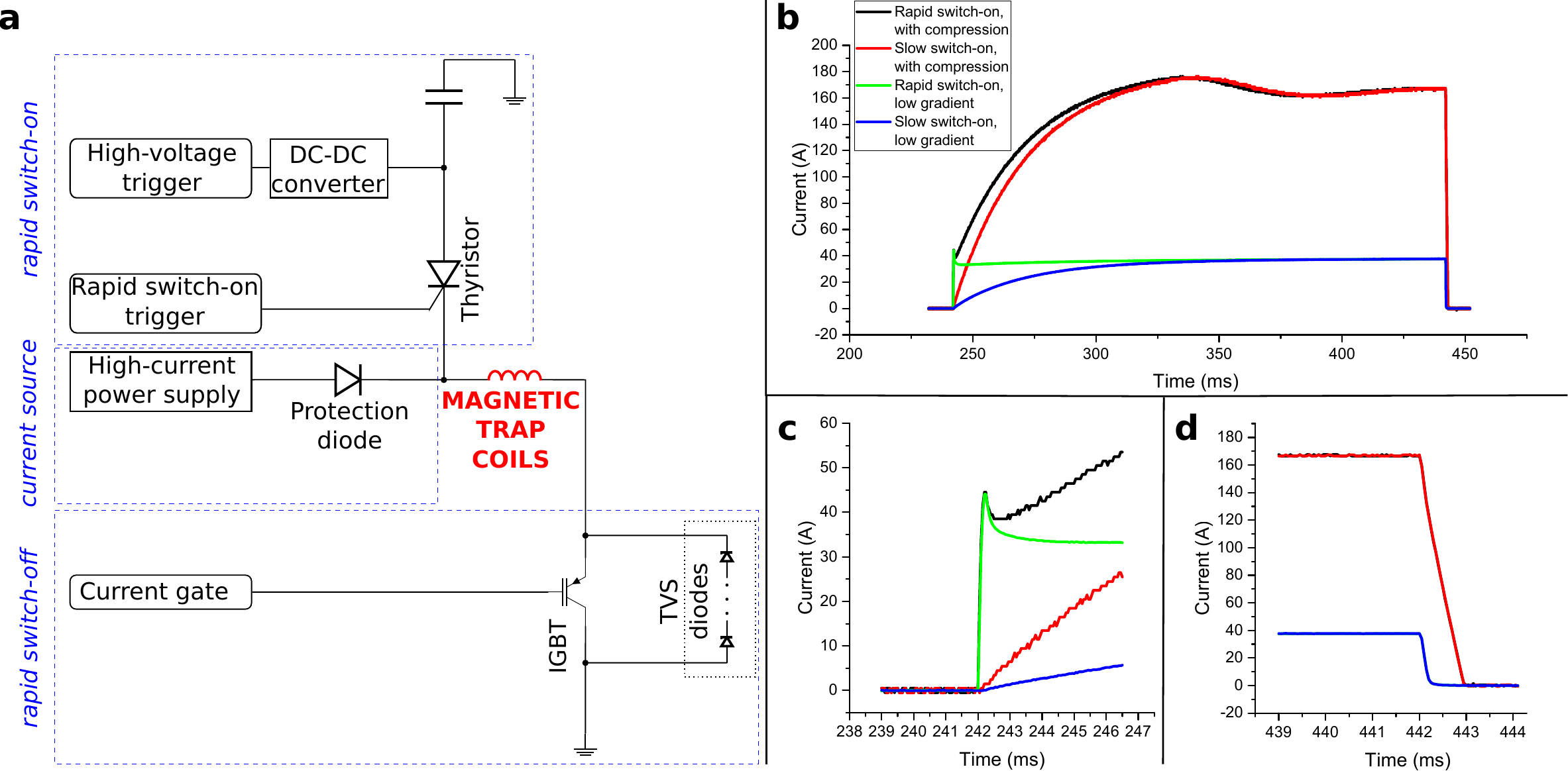}
    \caption{(color online) Rapid current-switching circuit. \textbf{a} Circuit diagram. \textbf{b} Measured current in trapping coils vs. time. Sample waveforms are shown for cases with and without rapid switch-on, as well as cases with and without MQT compression to the full $165$~A current ($140$~G/cm gradient) supported by the power supply. \textbf{c--d} Detail of switch-on (c) and switch-off (d) performance for each of these cases.}
    \label{fastswitch}
\end{figure*}

The magnetic trapping coils consist of $14$ layers of $8$ turns each, with inner diameter diameter $\approx\!7$~in and outer diameter $\approx\!9$~in, and have a mean vertical separation of $\approx\!6$~in. 
The $7/32\times7/32$~in$^2$ square wire is water-cooled, through a small hole in the center of the wire. The two coils are wired in series, with resistance $\approx\!0.2$~$\Omega$ and inductance $\approx\!3.6$~mH.

Due to the substantial inductance of our magnetic trapping coils, it is technically challenging to produce rapid changes in the current and therefore gradient applied in the trap. This rapid-switching capability is, however, highly desirable for magnetic trapping. In order to capture as many molecules as possible, it is important that the trapping coils be able to rapidly switch on to produce a gradient large enough to levitate molecules in the trapping state against gravity. In order to obtain accurate time-of-flight imaging data, it is important that the coils switch off rapidly to allow free evolution of the molecular cloud. To achieve switch-on and switch-off times under $1$~ms, we implement custom circuitry largely based on the designs described in Ref. [35].

The rapid switching circuitry can be roughly understood as consisting of three functional blocks (Fig. \ref{fastswitch}a). The first is the rapid switch-on block. This block contains a high-voltage DC-DC converter which charges a large, $10$~$\mu$F capacitor to up to $1$~kV (depending on the switch-on current desired). To trigger the rapid switch-on, a digital signal closes a thyristor and allows the capacitor to discharge into the trapping coils. This circuit allows magnetic field gradients of $32$~G/cm to be produced by our trapping coils within $250$~$\mu$s. (The rapid switch-on produces a transient $\approx\!20$\% overshoot of the target current, as can be seen in Fig. \ref{fastswitch}c.) The second functional block is the main, high-current power supply, which allows current to be maintained at a nearly constant level during magnetic trapping or to be increased in a ramp for adiabatic compression, up to $165$~A (corresponding to a magnetic field gradient of $140$~G/cm). This power supply is protected from the high voltages present during switch-on and switch-off by a high-current diode. The final functional block is the rapid switch-off circuitry, which includes a high-power insulated-gate bipolar transistor (IGBT) module and a series of transient voltage suppression (TVS) diodes. The IGBT module controls all current flow in the coils; when the IGBT module is closed, current may flow, and when the IGBT module is opened, current rapidly stops flowing. Due to the large inductance of the trapping coils, rapid switch-off of the current can cause very high voltages across the IGBT module. To limit these voltages to safe levels (here, $\sim600$~V), TVS diodes provide an alternate path to ground. These are arranged as a series of 11 pairs of TVS diodes. With this limiting voltage, switch-off from the full $165$~A current can be achieved in less than $1$~ms (Fig. \ref{fastswitch}d).

During testing, our CCD imaging camera was found to be sensitive to the magnetic fields produced at its original location very near the trapping coils. For this work, using a one-to-one telescope, the camera was moved far enough away from the trapping coils to reduce the effect from magnetic fields to a negligible level. This change was found to reduce the collection efficiency of our imaging optics by a factor of $1.8$ compared to previous work and to reduce the field of view to 17.8 mm by 13.3 mm (compared to 19.9 mm by 14.9 mm previously) [15].

\section*{Adiabaticity of MQT compression}
In order for compression to be adiabatic, we require $d\omega/dt \ll \omega^2$, where $\omega$ is the frequency of molecular motion in the trap. For the linear potential in an MQT, there is not a well-defined trap frequency, so we identify an effective trap frequency $\omega^\text{eff}\sim v/r$, 
where $v$ is the velocity of a molecule in the trap and $r$ is the radius of its orbit. For the most strongly-trapped states, using the virial theorem, we have $m v^2 \sim \mu_B G r$, where $m=107$~amu is the mass of SrF, $\mu_B$ is the Bohr magneton, and $G$ is the magnetic field gradient. From the equipartition theorem, we have $m v^2 \sim k_B T$, where $k_B$ is the Boltzmann constant and $T$ is the temperature. 
We have the initial effective trap frequency $\omega_0^\text{eff} \sim \mu_B G_0 / (\sqrt{m k_B T_0})$, where $T_0$ is the expected initial temperature in the trap and $G_0$ the initial trap gradient. We take $G_0$ to be the geometric mean of the switch-on gradient in the axial and radial dimensions. If the compression is adiabatic, we would expect $r \propto G^{-1/3}$ and $v \propto G^{1/3}$ [38], so that $\omega^\text{eff} \propto G^{2/3}$. Then $\omega^\text{eff}(t) = \omega_0^\text{eff} (G(t)/G_0)^{2/3}$, and 
\[\frac{d\omega^\text{eff}}{dt} = \frac{2\omega_0^\text{eff}}{3 G_0^{2/3} G^{1/3}} \frac{dG}{dt}.\] 
Under our conditions, $T_0 \approx90$~$\mu$K and $G_0 \approx 20$~G/cm, so that $\omega_0^\text{eff} \sim \mu_B G_0/(\sqrt{m k_B T_0}) \sim 100$~Hz. With our standard compression timing, the gradient doubles in the first $13$~ms after switch-on of the trap, and we have
\[ \frac{d\omega^\text{eff}}{dt} \sim \frac{2\omega_0^\text{eff}}{3 G_0} \frac{2 G_0}{13\text{ ms}} \sim \left(100\text{ Hz}\right)\cdot \omega_0^\text{eff}\sim \left(\omega^\text{eff}_0\right)^2.\]
This is inconsistent with the adiabatic condition that was assumed. Hence, it is plausible that the MQT compression may not be well described by the usual adiabatic compression formulation. This means that the phase-space density of molecules may decrease during the early part of the compression ramp.  Later in the ramp, $dG/dt$ remains the same or decreases, while $\omega^{\text{eff}}$ increases, so adiabaticity should be increasingly satisfied.

\end{document}